\documentclass[11pt,a4paper]{emulateapj}
\usepackage{graphicx}
\usepackage{textcomp}
\usepackage{amsmath}
\usepackage{amssymb}
\usepackage{gensymb}
\usepackage{rotating}
\usepackage{epsfig}
\usepackage{amsmath}
\usepackage{longtable}
\usepackage{color}
 \def\apj{ApJ}
 \def\apjl{ApJL}
 \def\mnras{MNRAS}
 \def\pasp{PASP}

 \def\aap{A\&A}
 
 \def\apjs{ApJS}
 
 \def\nat{Nature}

 \def\gs{\mathrel{\raise0.35ex\hbox{$\scriptstyle >$}\kern-0.6em\lower0.40ex\hbox{{$\scriptstyle \sim$}}}}
 \def\ls{\mathrel{\raise0.35ex\hbox{$\scriptstyle <$}\kern-0.6em\lower0.40ex\hbox{{$\scriptstyle \sim$}}}}

 \def\Msol{\mathrel{\rm M_{\odot}}}
 \def\Lsol{\mathrel{\rm L_{\odot}}}
 \def\Msolyr{\mathrel{\rm M_{\odot}\,yr^{-1}}}
 \def\Wm2{\,\hbox{W}\,\hbox{m}^{-2}}
 \def\gsim{\mathrel{\raise0.35ex\hbox{$\scriptstyle >$}\kern-0.6em\lower0.40ex\hbox{{$\scriptstyle \sim$}}}}
 \def\lsim{\mathrel{\raise0.35ex\hbox{$\scriptstyle <$}\kern-0.6em\lower0.40ex\hbox{{$\scriptstyle \sim$}}}}
 
 \def\pc{\%}

\lefthead{Simpson et al.}  
\righthead{ZF20115: a ``passive'' starburst galaxy?}

\begin{document}

\title{An imperfectly passive nature: Bright sub-millimeter emission from dust-obscured star formation in the $z$\,=\,3.717 ``passive'' system, ZF\,20115}

\author{
J.\,M.\ Simpson,\altaffilmark{1}
Ian Smail,\altaffilmark{2}
Wei-Hao Wang,\altaffilmark{1}
D.\ Riechers,\altaffilmark{3} 
J.\,S.\ Dunlop,\altaffilmark{4}
Y.\,Ao,\altaffilmark{5}
N.\,Bourne,\altaffilmark{4}
A.\ Bunker,\altaffilmark{6}
S.\,C.\ Chapman,\altaffilmark{7}
Chian-Chou\ Chen,\altaffilmark{8}
H.\ Dannerbauer,\altaffilmark{9,10}
J.\,E.\ Geach,\altaffilmark{11}
T.\ Goto,\altaffilmark{12}
C.\,M.\ Harrison,\altaffilmark{8}
H.\,S.\ Hwang,\altaffilmark{13}
R.\,J.\ Ivison,\altaffilmark{8,4}
Tadayuki Kodama,\altaffilmark{5}
C.\,-H.\ Lee,\altaffilmark{14}
H.\,-M.\ Lee,\altaffilmark{15}
M.\ Lee,\altaffilmark{5,16}
C.\,-F.\ Lim,\altaffilmark{1}
M.\,J.\ Micha{\l}owski,\altaffilmark{4}
D.\,J.\ Rosario,\altaffilmark{2}
H.\ Shim,\altaffilmark{17}
X.\,W.\ Shu,\altaffilmark{18}
A.\,M.\ Swinbank,\altaffilmark{2}
W.\,-L.\ Tee,\altaffilmark{1,19}
Y.\ Toba,\altaffilmark{1}
E.\ Valiante,\altaffilmark{20}
Junxian Wang,\altaffilmark{21}
X.~Z.~Zheng,\altaffilmark{22}
}

\setcounter{footnote}{0}
\altaffiltext{1}{EACOA fellow: Academia Sinica Institute of Astronomy and Astrophysics, No.\ 1, Sec.\ 4, Roosevelt Rd., Taipei 10617, Taiwan; email: jsimpson@asiaa.sinica.edu.tw} 
\altaffiltext{2}{Centre for Extragalactic Astronomy, Department of Physics, Durham University, South Road, Durham DH1 3LE, UK} 
\altaffiltext{3}{Department of Astronomy, Cornell University, 220 Space Sciences Building, Ithaca, NY 14853, USA}
\altaffiltext{4}{Institute for Astronomy, University of Edinburgh, Royal Observatory, Blackford HIll, Edinburgh EH9 3HJ, UK}
\altaffiltext{5}{National Astronomical Observatory of Japan (NAOJ), 2-21-1 Osawa, Mitaka, Tokyo 181-8588, Japan}
\altaffiltext{6}{Department of Physics, University of Oxford, Denys Wilkinson Building, Keble Road, Oxford, OX1 3RH, United Kingdom}
\altaffiltext{7}{Department of Physics and Atmospheric Science, Dalhousie University, Halifax, NS B3H 3J5 Canada}
\altaffiltext{8}{European Southern Observatory, Karl Schwarzschild Strasse 2, Garching, Germany}
\altaffiltext{9}{Instituto de Astrofi­sica de Canarias (IAC), E-38205 La Laguna, Tenerife, Spain }
\altaffiltext{10}{Universidad de La Laguna, Dpto. Astrofi­sica, E-38206 La Laguna, Tenerife, Spain}
\altaffiltext{11}{Centre for Astrophysics Research, Science and Technology Research Institute, University of Hertfordshire, Hatfield AL10 9AB, UK}
\altaffiltext{12}{Institute of Astronomy, National Tsing Hua University, No.\ 101, Section 2, Kuang-Fu Road, Hsinchu, Taiwan. 30013}
\altaffiltext{13}{School of Physics, Korea Institute for Advanced Study, 85 Hoegiro, Dongdaemun-gu, Seoul 02455, Korea}
\altaffiltext{14}{Subaru Telescope, National Astronomical Observatory of Japan 650 N Aohoku Pl., Hilo, HI 96720, USA}
\altaffiltext{15}{Department of Physics and Astronomy, Seoul National University, 1 Gwanak-ro, Gwanak-gu, Seoul 08826, Korea}
\altaffiltext{16}{Department of Astronomy, The University of Tokyo, 7-3-1 Hongo, Bunkyo-ku, Tokyo 133-0033, Japan}
\altaffiltext{17}{Department of Science Education, Kyungpook National University, 80 Daehakto, Bukgu, Daegu 41566, Korea}
\altaffiltext{18}{Department of Physics, Anhui Normal University, Wuhu, Anhui, 241000, China}
\altaffiltext{19}{Department of Physics, National Taiwan University, Taipei 10617, Taiwan}
\altaffiltext{20}{School of Physics and Astronomy, Cardiff University, Queen's Buildings, The Parade, Cardiff CF24 3AA, UK}
\altaffiltext{21}{Center for Astrophysics, University of Science \& Technology of China, Hefei, Anhui 230026, P.\ R.\ China}
\altaffiltext{22}{Purple Mountain Observatory, Chinese Academy of Sciences, 2 West Beijing Road, Nanjing 210008, China}

\begin{abstract}
The identification of high-redshift, massive galaxies with old stellar populations may pose challenges to some models of galaxy formation. However, to securely classify a galaxy as quiescent, it is necessary to exclude significant ongoing star formation, something that can be challenging to achieve at high redshifts. In this letter, we analyse deep ALMA\,/\,870\,$\mu$m and SCUBA-2\,/\,450\,$\mu$m imaging of the claimed ``post-starburst'' galaxy ZF-20115 at $z=$\,3.717 that exhibits a strong Balmer break and absorption lines. The far-infrared imaging reveals a luminous starburst located 0.4\,$\pm$\,0.1$''$ ($\sim$\,3\,kpc in projection) from the position of the rest-frame ultra-violet\,/\,optical emission, with an obscured star-formation rate of 100$^{+15}_{-70}$\,$\Msolyr$. This star-forming component is undetected in the restframe ultraviolet but contributes significantly to the lower angular resolution photometry at restframe wavelengths $\gsim$\,3500$\mathrm{\AA}$, significantly complicating the determination of a reliable stellar mass. Importantly, in the presence of dust obscuration, strong Balmer features are not a unique signature of a post-starburst galaxy and are indeed frequently observed in infrared-luminous galaxies. We conclude that the ZF\,20015 system does not pose a challenge to current models of galaxy formation and that deep sub-/millimeter observations are a pre-requisite for any claims of quiescence. The multi-wavelength observations of ZF\,20115 unveil a complex system with an intricate and spatially-varying star-formation history. ZF\,20115 demonstrates that understanding high-redshift obscured starbursts will only be possible with multi-wavelength studies that include high-resolution observations, available with the JWST, at mid-infrared wavelengths. 
\end{abstract}


\section{Introduction}
\label{sec:intro}
In the local Universe, the most massive galaxies are giant spheroidal galaxies that formed the bulk of their stellar populations in a burst of star formation at $z$\,$\gsim$\,2 (e.g.\ \citealt{Nelan05}). Identifying the progenitors of these galaxies, in either a high-redshift passive or starburst phase, has become a major focus of galaxy formation surveys (e.g.\ \citealt{Simpson14,Straatman14}). 

One route to isolate passive galaxies at high redshifts is to search for galaxies that have extremely red colors (e.g.\ $H$--4.5\,$\mu$m\,$>$\,4). These colors may arise due to the presence of a redshifted Balmer (3646\,$\mathrm{\AA}$) or 4000\,$\mathrm{\AA}$ break in the spectral energy distribution (SED) of the source. However, the degeneracy between stellar age, redshift and dust extinction means that proposed ``passive'' samples, selected on apparent colors, suffer high ($\gsim$\,80\,\%) contamination from dusty interlopers at low-- and high--redshift \citep{Smail02b,Toft05,Boone11,Caputi12}. Indeed, an early attempt to identify a $z$\,$>$\,3 passive galaxy was presented by \citet{Mobasher05}, who claimed the detection of a bright post-starburst galaxy at $z$\,$\sim$\,6.5. \citet{Dunlop07} subsequently showed that allowing for extreme values of dust obscuration in the SED model yields a high likelihood that the source is a $z$\,$\sim$\,2 dusty starburst. 

Despite these challenges a number of studies have continued to claim the detection of large numbers of passive galaxies at high redshift from wide-field near-infrared imaging (\citealt{Marchesini10,Nayyeri14}). The most massive of these are thought to have stellar masses of $\gsim$\,10\,$^{11}$\,$\Msol$ at $z\gsim$\,3 that, if correct, may pose challenges for models of galaxy formation. However a convincing spectroscopic confirmation of a truly massive, quiescent galaxy at $z\gsim$\,3 has yet to be presented. 

Recently, \citet{Glazebrook17}  presented deep near-infrared spectroscopy of ZF\,20115, a purported ``passive'' galaxy at $z$\,=\,3.717. ZF\,20115 was selected from the ZFOURGE survey based on the presence of a strong Balmer break identified in near-infrared photometry \citep{Straatman14}. Near-infrared spectroscopy then confirmed Balmer absorption lines with high equivalent width (EW) that were suggested to show that ZF\,20115 is a ``post-starburst'' galaxy with a stellar age of 0.2--1\,Gyr. The Balmer lines, combined with fits to the broad-band photometry, led \citet{Glazebrook17} to conclude in favour of an age in the range 0.5--1 Gyr, corresponding to a formation redshift $z_{\mathrm{form}}\sim$\,5--8. Combined with the estimated stellar mass of 1.5--1.8\,$\times$\,10$^{11}$\,$\Msol$ this indicates a rapid conversion of baryons into stars at high redshift, as expected from studies of sub-millimeter galaxies (SMGs; e.g.\ \citealt{Lilly99, Smail02}).

In this letter we analyse sub-millimeter observations of ZF\,20115 with SCUBA-2 and ALMA which identify an intense, obscured starburst within 0.4\,$\pm$\,0.1\,$''$ of the restframe ultraviolet component (ZF\,20115--UV). Throughout we adopt a $\Lambda$CDM cosmology with $H_{\rm 0}$\,=\,70\,km\,s$^{-1}$\,Mpc$^{-1}$, $\Omega_{\Lambda}$\,=\,0.7, and $\Omega_{\rm m}$\,=\,0.3 and a Chabrier initial mass function \citep{Chabrier03}.

\section{Observations}
The galaxy ZF\,20115, located in the CANDELS region of the COSMOS field, was selected based on its rest-frame optical color in the ZFOURGE survey and we use this photometric catalogue in our analysis \citep{Straatman16}.

ZF\,20115 was observed in ALMA Cycle\,2 at 870\,$\mu$m for 1.4\,min as part of program 2013.1.01292.S. The ALMA observations employed 39 antennae yielding a synthesized beam of 1.1$''$\,$\times$\,0.6$''$, reach a depth of $\sigma_{870}$\,=\,0.2\,mJy\,beam$^{-1}$ and reveal a significant (6.9\,$\sigma$) source (ZF\,20115--FIR) within 0.4\,$\pm$\,0.1$''$ of ZF\,20115--UV. We use {\sc CASA}\,/\,{\sc imfit} to model the emission and determine that ZF\,20115--FIR is unresolved with a total flux density of 1.4\,$\pm$\,0.2\,mJy\,beam$^{-1}$.

The central area of the CANDELS\,/\,COSMOS is being mapped at 450 and 850\,$\mu$m by the SCUBA-2 Ultra Deep Imaging EAO Survey (STUDIES: PI: W.-H.\ Wang), a large program at the James Clerk Maxwell Telescope (JCMT). We make use of the first STUDIES release, which reaches a depth of $\sigma_{450}$\,$\sim$\,1\,mJy in the vicinity of the ZF\,20115\,\footnote{Including archival data from \citet{Geach17}.}. We detect ZF\,20115--FIR at a 3\,$\sigma$ significance level in the 450\,$\mu$m imaging, identifying a $S_{450}$\,=\,3.1\,$\pm$\,1.0\,mJy source within the expected 1--\,$\sigma$ positional uncertainty (2.1$''$; \citealt{Ivison07}). ZF\,20115--FIR is detected in the STUDIES SCUBA-2 850\,$\mu$m imaging with a flux density of $S_{850}$\,=\,1.49\,$\pm$\,0.15\,mJy, consistent with the ALMA detection.

The CANDELS\,/\,COSMOS region was imaged with {\it Herschel}\,/\,PACS and SPIRE as part of the {\it Herschel}--CANDELS survey and {\it Herschel} Multi-tiered Extragalactic Survey (HerMES; \citealt{Oliver12}). \citet{Straatman14} present the PACS photometry for ZF\,20115, with stated 1--$\sigma$ uncertainties of 0.4\,mJy at 100 and 160\,$\mu$m. At 160\,$\mu$m the claimed depth is below the measured confusion limit of PACS but it is not possible to verify this as the reduced {\it Herschel}--CANDELS imaging is not publicly available and the data reduction is not detailed in the literature. The deblended SPIRE imaging reaches 1--$\sigma$ depths of 3.1, 3.5 and 4.1\,mJy at 250, 350, and 500\,$\mu$m \citep{Swinbank14}. ZF\,20115 is not detected in the 100--500\,$\mu$m imaging. Finally, ZF\,20115 is not detected in the available 1.4 and 3\,GHz imaging \citep{Schinnerer10,Smolcic17}, consistent with lying at a redshift of $z$\,$\gsim$\,2.5.

We align the astrometry of the wide-field imaging presented here by correcting for the mean offset between the {\it Spitzer} IRAC\,/\,3.6\,$\mu$m image and the relevant image. We cannot apply this technique to the single ALMA pointing but we note that the overall ALMA astrometry has been found to be in agreement with the solution for the CANDELS\,/\,COSMOS imaging \citep{Schreiber17}. 

%
%
\begin{figure}
  \centering
  \includegraphics[width=0.9\columnwidth]{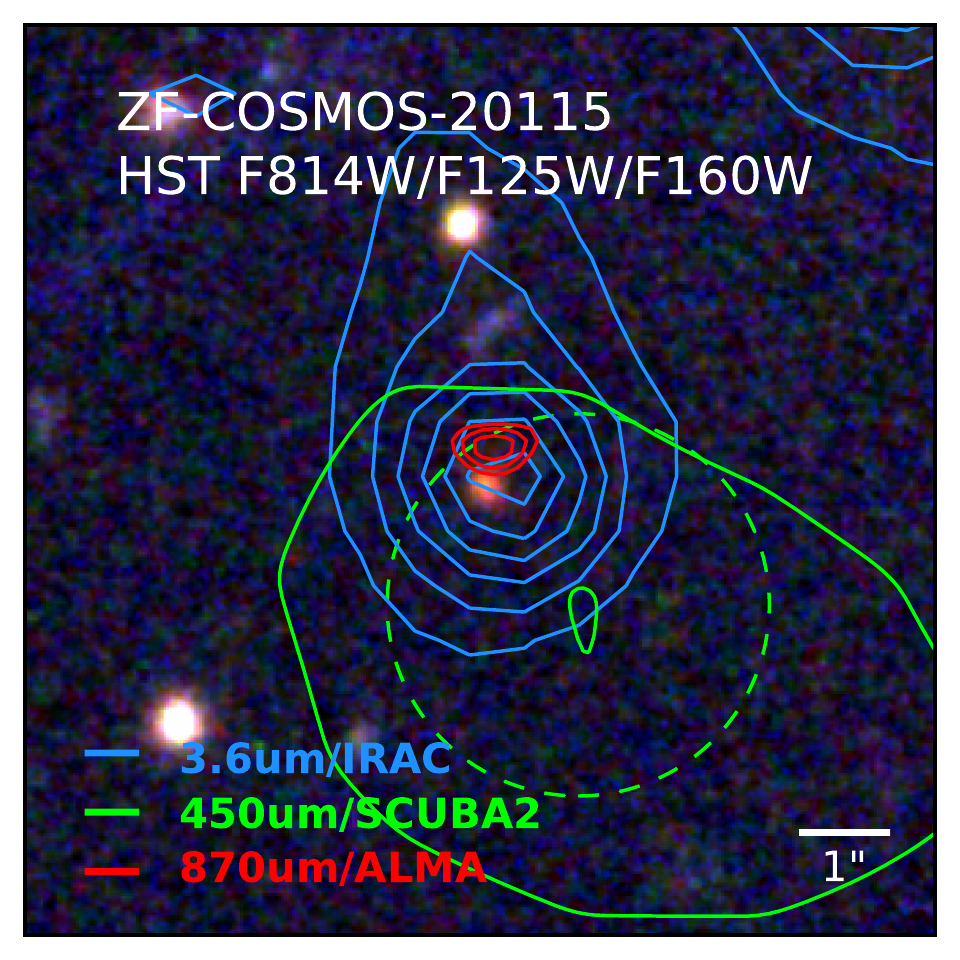} 
  \caption{HST true color image ($I_{814}J_{125}H_{160}$) of ZF\,20115 overlaid with contours representing: ALMA\,/\,870\,$\mu$m ($\pm$\,4,\,5 ...\,$\times$\,$\sigma$); SCUBA-2\,/\,450\,$\mu$m ($\pm$\,2,\,3\,$\times$\,$\sigma$); and IRAC\,/\,3.6\,$\mu$m ($\pm$\,5,\,10,\,15 ...\,$\times$\,$\sigma$) emission . The 1-$\sigma$ positional uncertainty on the 450\,$\mu$m detection is represented by a dashed circle. The ALMA emission is offset by 0.4\,$\pm$\,0.1$''$ ($\sim$\,3\,kpc in projection) to the rest-frame ultraviolet--to--optical emission, consistent with {\it HST} observations of high-redshift dusty starbursts \citep{Chen15}. The 3.6\,$\mu$m\,/\,IRAC emission appears extended in the direction of the far-infrared emission, highlighting that the near-infrared photometry of ZF\,20115 likely comprises a blend of the unobscured and obscured components. 
}
\label{fig:hst_submm}
\end{figure}

\section{Analysis}

In Figure\,1 we show the archival ALMA 870\,$\mu$m data for ZF\,20115, contoured on the {\it HST}\,/\,CANDELS imaging. We first consider that the observed 870\,$\mu$m emission (ZF\,20115--FIR) is located 0.4\,$\pm$\,0.1$''$ ($\sim$\,3\,kpc in projection) from the detected rest-frame $UV$ emission. At the redshift of ZF\,20115--UV ($z$\,=\,3.717) the near-infrared wavebands ($\lsim$\,2.0\,$\mu$m) trace rest-frame ultraviolet emission and thus it is unsurprising to find positional offsets between obscured and unobscured components. Indeed, the measured offset between ZF\,20115--UV and ZF\,20115--FIR is consistent with \citet{Chen15}, who determine an {\it intrinsic} offset between the {\it HST}- and ALMA-traced emission in SMGs of 0.55$''$ (1\,$\sigma$), evidence for structured dust obscuration in these sources \citep{Chapman04b,Hodge15}.

To test whether ZF\,20115--UV and ZF\,20115--FIR are physically associated we calculate the probability that they arise due to the chance alignment of two sources on the sky. First, we create mock catalogues with source surface densities that satisfy the ZFOURGE color-criterion for a massive, quiescent galaxy \citep{Straatman14}, and sub-millimeter sources at $S_{870}>$\,1\,mJy \citep{Dunlop17}. We do not account for any bias due to gravitational lensing but comment that this is unlikely given that ZF\,20115 is located at $z$\,=\,3.717. From these mock catalogues we determine a probability of $\sim$\,1.5\,$\times$\,10$^{-4}$ that ZF\,20115--UV and ZF\,20115--FIR are a chance alignment, in agreement with the corrected Poissonian probability (\citealt{Downes86}). Given this low probability, and the similarity with the properties of high-redshift dusty starbursts, we conclude that ZF\,20115 is a composite dusty starburst.

\subsection{Properties of the starburst}

To determine the far-infrared properties of ZF\,20115--FIR we fit the far-infrared photometry with a single-temperature, optically-thin, modified blackbody (mBB) function. The 100\,$\mu$m photometry (restframe $\sim$\,20\,$\mu$m) is dominated by Polycylic Aromatic Hydrocarbon features and is not included in our SED fitting. The dust emissivity is fixed at $\beta=$\,1.8 and the best-fit parameters and associated uncertainties are determined using a Monte-Carlo Markov Chain (MCMC) approach \citep{Simpson17}, accounting for the effect of the Cosmic Microwave Background following \citet{dacuhna12}.

Our SED fitting shows that ZF\,20115--FIR has a best-fit dust temperature of $T_{\mathrm{d}}$\,=\,30.4$^{+1.4}_{-12.0}$\,K and a far-infrared luminosity (8--1000\,$\mu$m) of $L_{\mathrm{FIR}}$\,=\,7.7$^{+1.0}_{-5.2}$\,$\times$\,10$^{11}$\,$\Lsol$. A single temperature mBB is known to underestimate the total far-infrared luminosity by $\sim$\,20\,\% (\citealt{Swinbank14}) and correcting for this we determine that ZF\,20115 has a total far-infrared luminosity $L_{\mathrm{FIR}}$\,=\,9.2$^{+1.2}_{-6.2}$\,$\times$\,10$^{11}$\,$\Lsol$ and an obscured star-formation rate (SFR) of 100$^{+15}_{-70}$\,$\Msolyr$ \citep{Kennicutt98}.

\section{Discussion}

Interest in ZF\,20115 (e.g.\ \citealt{Rong17}) has arisen due to the claim that its stellar population is massive and quiescent at $z$\,=\,3.717 \citep{Glazebrook17}. The presence of an obscured starburst, blended or co-located, with the optical-to-near-infrared emission requires a re-assessment of the properties of ZF\,20115.

\subsection{Origin of Balmer Absorption Lines}\label{subsec:balmer}

Near-infrared spectroscopy of ZF\,20115 identified Balmer absorption lines with a combined EW of 38\,$\pm$\,6\,$\mathrm{\AA}$ (summing H$\beta$, H$\gamma$, and H$\delta$; \citealt{Glazebrook17}). This requires that the detectable stellar continuum around 4000--5000$\mathrm{\AA}$ is dominated by A-stars with lifetimes of $\sim$\,0.1--1\,Gyr and is often taken as an indicator of a ``post-starburst'' population. However, Balmer absorption features are not unique to post-starburst galaxies; they appear frequently in dusty starburst galaxies as well \citep{Poggianti00}.

\citet{Poggianti00} present detections of the H\,$\delta$ absorption line for a sample of {\it IRAS}--selected galaxies ($L_{\mathrm{FIR}}$\,$>$\,5\,$\times$\,10$^{11}$\,$\Lsol$) at $z$\,$\lsim$\,0.05. This far-infared-bright sample is dominated by star-forming galaxies, and over 60\,\% of the sources show evidence of on-going interactions or mergers. In Figure~2 we show the SED of Mrk\,331 \citep{Brown14} from the sample of \citet{Poggianti00}, scaled in luminosity to match ZF\,20115. Mrk\,331 is a late-stage merger and exhibits H$\delta$ absorption with EW\,=\,4.1$\mathrm{\AA}$ and a strong break at $\lsim$\,4000\,$\mathrm{\AA}$. We note that few local examples reach the high EW Balmer absorption observed from ZF\,20115 but comment that this may be a reflection of the low survey volumes and decreased activity at $z$\,$\sim$\,0. The SED of Mrk\,331 is in qualitative agreement with that of ZF\,20115, illustrating the challenge of deriving accurate star-formation histories for high-redshift ``red'' galaxies from optical-to-near-infrared photometry alone. 

The high EW Balmer absorption features that appear common in starburst galaxies can be explained by age-dependant dust obscuration \citep{Poggianti00}. This can be considered in terms of a single isolated galaxy, or triggered star-formation in an ongoing merger. Given the presence of two components with a projected offset of 3\,kpc we suggest that the ZF\,20115 system is an ongoing merger between two gas-rich progenitors. In this scenario the centroid of the {\it observable} stellar emission likely traces one component of the merger that underwent a burst of star formation, triggered by an earlier interaction.

\citet{Hopkins13} present numerical simulations of the merger of gas-rich disks at high-redshift, which demonstrate that star-formation is triggered in the individual disks $\sim$\,0.5\,Gyr before coalescence, and the triggering of a nuclear starburst. The Balmer absorption in the spectrum of ZF\,20115 is likely the consequence of a similar merger event, with the strength of the lines indicating a minimum age for this initial star-formation episode of $\sim$\,100\,Myr \citep{Delgado99,Glazebrook17}.

%
%
\begin{figure}
  \centering
  \includegraphics[width=0.45\textwidth]{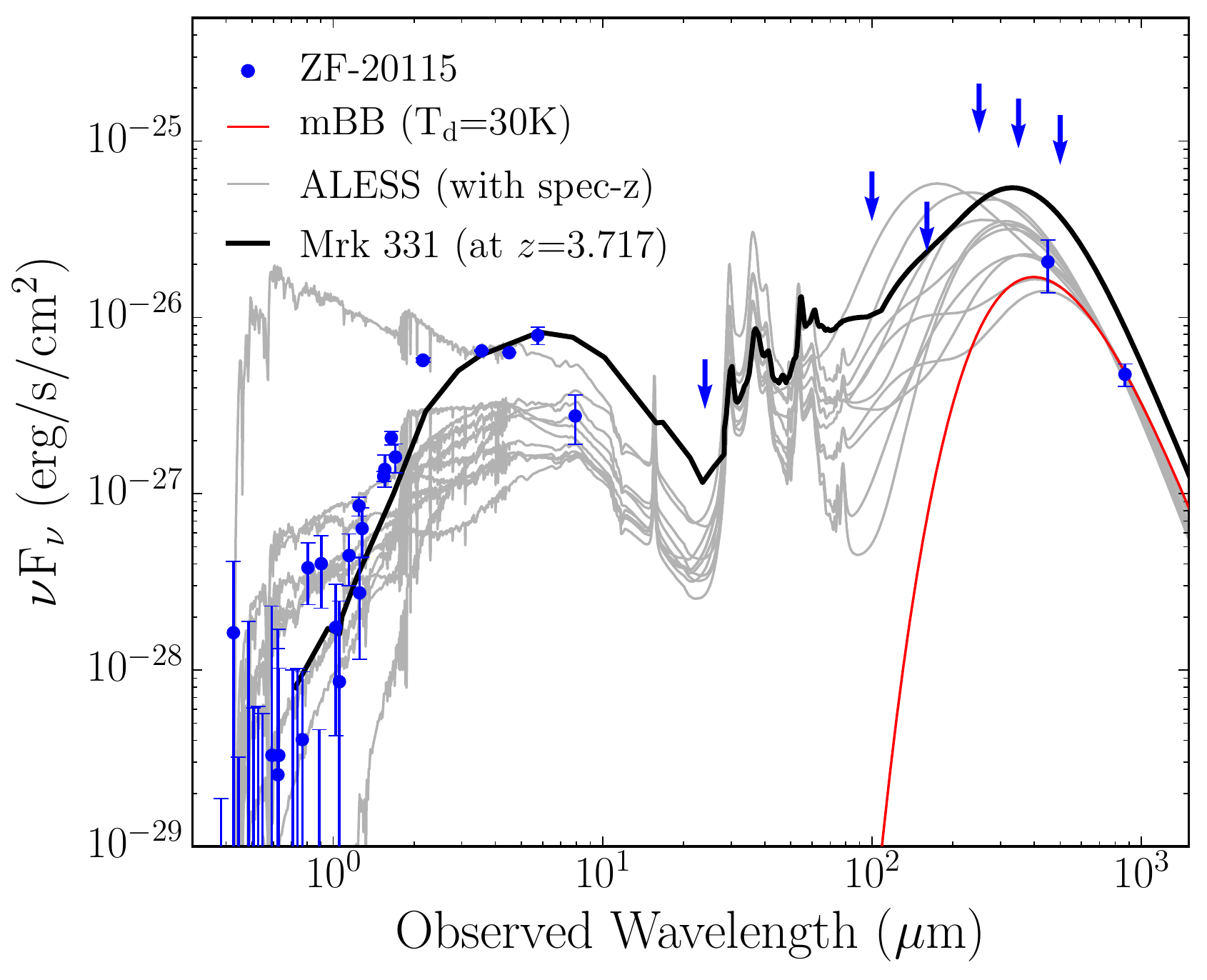}
  \caption{The observed photometry of ZF\,20115 as a function of observed wavelength (arrows represent 3\,$\sigma$ upper limits). The best-fit mBB function to the far-infrared photometry is shown and corresponds to an obscured SFR of 100$^{+15}_{-70}$\,$\Msolyr$. We overlay SEDs derived for a subset of the spectroscopically-confirmed ALESS SMGs \citep{Danielson17}, which indicate the significant contribution that the obscured starburst makes to the overall optical-to-near-infrared photometry (20--75\,\%). The model SED of Mrk\,331 is shown, redshifted to $z=$\,3.717 and scaled to broadly match ZF\,20115. This local far-infrared--bright starburst is an ongoing merger, has detected H$\beta$ and H$\delta$ in absorption, and a strong Balmer break. The photometry of Mrk\,331 is in reasonable agreement with ZF\,20115, illustrating that Balmer features do not uniquely identify ``post-starburst'' galaxies.
}
\label{fig:sed}
\end{figure}

\subsection{Stellar mass of ZF\,20115}

The detection by ALMA of a strongly star-forming, obscured component within 0.4\,$\pm$\,0.1$''$ of the claimed quiescent component ZF\,20115--UV significantly complicates the analysis of this system. To some extent the issue about whether the two components are part of a single ``galaxy'' or are distinct components within a larger halo, is one of semantics. What is unarguable is that the presence of this luminous, star-forming component will result in significant contamination of the flux measurements in longer wavelength bands, which are critical for estimating the stellar mass of the system.  At the redshift of ZF\,20115, the stellar emission from the obscured starburst peaks in the 2$''$--resolution {\it Spitzer}\,/\,IRAC imaging (rest-frame 0.8--1.7\,$\mu$m), where the two components are strongly blended (Figure\,1). At these wavelengths, the mass-to-light ratio of a stellar population varies by an order of magnitude for ages of 50--1000\,Myr \citep{Hainline11} and so it is crucial that we take the contribution from ZF\,20115--UV into account. 

We use two methods to estimate the likely contribution from ZF\,20115--FIR to the integrated photometry. First, we deblend the 3.6\,/\,4.5\,$\mu$m emission using three 2-D Gaussian components representing ZF\,20115--UV, ZF\,20115--FIR, and the northern optically-bright galaxy (Figure\,1). Using an MCMC approach we determine that ZF20115–FIR contributes $\sim$\,40-–50$\pc$ of the integrated flux density at 3.6 and 4.5\,$\mu$m, consistent with the two components being indistinguishable in the IRAC imaging. Second, we estimate the contribution by considering the spectroscopically-identified, ALMA-located--SMGs from \citet{Danielson17}. We fit the templates for these SMGs to the far-infrared photometry of ZF\,20115--FIR and in Figure\,2 show those SEDs that are near-infrared--bright and have far-infrared properties matching ZF\,20115--FIR. We create an average SED from these best-fit templates and estimate that the obscured starburst contributes 20--75\,\% (median 33\,$\pm$\,4\,\%) of the measured photometry of ZF\,20115 at observed 1.5--8.0\,$\mu$m, broadly consistent with our estimates from deblending the IRAC imaging. 

To estimate the mass of the stellar component associated with ZF\,20115--UV we subtract the average starburst SED from the global photometry. Using {\sc magphys} \citep{dacunha15} to model the residual photometry we estimate that this component has a stellar mass of $\sim$\,0.8\,$\times$\,10$^{11}$\,$\Msol$, less than half of that estimated by \citet{Glazebrook17}. We cannot determine an accurate star-formation history for this component but note that a rapid formation at high redshift ($z$\,$\gsim$\,5) is no longer required and that the implied SFR and duration are consistent with the population of obscured starbursts that are known to reside at $z$\,$\gsim$\,4 (e.g.\ \citealt{Riechers13,Strandet16,Simpson17}). Using the stellar mass estimates presented by \citet{Danielson17} we estimate that ZF\,20115--FIR has a mass of $\sim$\,0.4\,$\times$\,10$^{11}$\,$\Msol$. At the current SFR and assuming a constant star-formation history this stellar component could form in $\sim$\,350\,Myr, equivalent to a formation redshift of $z$\,=\,3.7--4.6. 

The apparent quiescent nature of ZF\,20115 led \citet{Glazebrook17} to suggest that this source, and the parent sample of proposed quiescent galaxies, may be in conflict with current models of galaxy formation (e.g.\ \citealt{Steinhardt16}). The contamination of the stellar mass estimates for ZF\,20115--UV has implications for the conclusions presented by \citet{Glazebrook17} and we revisit those here. 

First, we adopt the halo mass of $\sim$\,3\,$\times$\,10$^{12}$\,$\Msol$ for ZF\,20115 estimated by \citet{Glazebrook17}. Assuming a cosmic baryon fraction of 16\% \citep{Planck16}, we estimate that the stellar mass for ZF\,20115--UV corresponds to a baryon conversion efficiency of $\sim$\,15\,--40\,$\pc$, at a formation redshift of $z_{\mathrm{form}}$\,=\,4--5. \citet{Genel14} demonstrate that comparable halos in the Illustris simulation have a conversion efficiency of $\sim$10\,$\pc$ at $z$\,$\sim$4, consistent with observational studies of $UV$--selected galaxies (\citealt{Finkelstein15}). The conversion of baryons into stars appears to be progressing more efficiently in ZF\,20115--UV, although it remains well below the theoretical limit. Indeed, simulations of galaxy formation do predict the presence of $z$\,$\sim$\,4 star-forming systems with stellar masses of $\sim$\,10$^{11}$\,$\Msol$ (\citealt{Wellons15,Gomez16}), at a space density in agreement with observational studies (e.g.\ \citealt{Mortlock17}). Thus, while ZF\,20115 remains an interesting system it does not appear to be in conflict with our understanding of galaxy formation.

\section{Conclusions} 
In this letter we have analysed ALMA and SCUBA-2 detections of far-infrared emission from ZF\,20115, a proposed massive, quiescent galaxy at $z$\,=\,3.717. The far-infrared imaging locates an obscured starburst only 0.4\,$\pm$\,0.1$''$ from the unobscured ZF\,20115--UV. The far-infrared luminous starburst appears associated with ZF\,20115--UV bringing into doubt the claim that this is a quiescent system. Indeed, we show that the Balmer absorption features exhibited by ZF\,20115 are not a unique tracer of a ``post-starburst'' galaxy, such features appear frequently in local {\it IRAS}--selected starbursts \citep{Poggianti00}. 

ZF\,20115--UV and ZF\,20115--FIR are separated by $\sim$\,3\,kpc on the sky, a strong indication that they reside within the same halo at $z$\,=\,3.717. Indeed, {\it HST} studies of ALMA-identified SMGs identify similar offsets between the obscured and unobscured components of comparably luminous galaxies at these redshifts. What is unavoidable is that the presence of a starburst within 0.4\,$\pm$\,0.1$''$ of ZF\,20115--UV means that an allowance must be made for blending in the low-spatial resolution SED of this system, particularly at long wavelengths. 

Including a dusty starburst component in the SED fitting of ZF\,20115 indicates that $\sim$\,30--50\,$\pc$ of the total rest-frame 1--2\,$\mu$m emission arises from the on-going starburst. Correcting for this contribution and refitting the SED we determine that ZF\,20115--UV has a stellar mass of $\sim$\,0.8\,$\times$\,10$^{11}$\,$\Msol$, although we stress that the uncertainties on this measurement are large. The corrected stellar mass is $\sim$\,50\,$\pc$ lower than previously published values and as we show this removes the tension between ZF\,20115 and models of galaxy formation.

The multi-wavelength imaging and spectroscopy of the ZF\,20115 system provides key insights into the properties of obscured starbursts at high redshifts. ZF\,20115 comprises a blend of obscured and unobscured components that host intense but, spatially-variable, star-formation. The strength of the Balmer absorption features suggest that the system has a complex star-formation history and was undergoing a star-formation event at least 100\,Myrs prior to the current episode. This can be explained by adopting a merger-model for the system but is challenging to understand if secular processes dominate the formation mechanism of high-redshift starbursts. What is clear from our analysis is that to untangle this complex system, and thus truly understand high-redshift starbursts, we require the high-resolution imaging at mid-infrared wavelengths that will only be available with the James Webb Space Telescope.

\section*{Acknowledgements}
The authors thank Nick Scoville for sharing his catalog of ALMA observations and Alice Shapley for helpful comments. JMS acknowledges an EACOA fellowship. IRS acknowledges support from DUSTYGAL\,321334, a RS/Wolfson Merit Award and STFC (ST/L00075X/1). DR acknowledges support under grant number AST-1614213.

This paper makes use of data taken with SCUBA-2\,/\,JCMT, and ALMA (ADS/JAO.ALMA$\#$2013.1.01292.S). The JCMT is operated by the EAO on behalf of The National Astronomical Observatory of Japan, Academia Sinica Institute of Astronomy and Astrophysics, the Korea Astronomy and Space Science Institute, the National Astronomical Observatories of China and the Chinese Academy of Sciences, with additional support from the Science and Technology Facilities Council of the United Kingdom and participating universities in the United Kingdom and Canada.

\end{document}